\documentclass[sn-mathphys-num]{sn-jnl}

\usepackage{graphicx}%
\usepackage{multirow}%
\usepackage{amsmath,amssymb,amsfonts}%
\usepackage{amsthm}%
\usepackage{mathrsfs}%
\usepackage[title]{appendix}%
\usepackage{xcolor}%
\usepackage{textcomp}%
\usepackage{manyfoot}%
\usepackage{booktabs}%
\setcitestyle{numbers}
\usepackage{algorithm}%
\usepackage{algorithmicx}%
\usepackage{algpseudocode}%
\usepackage{listings}%
\usepackage{array}
\usepackage{mathrsfs}
\usepackage{mdwmath}
\usepackage{changepage}
\usepackage{algorithm}  
\usepackage{cuted}
\usepackage{mathbbol}
\usepackage{mdwtab}
\usepackage{eqparbox}
\usepackage{algpseudocode,float}
\usepackage{threeparttable}
\usepackage{lipsum}
\usepackage{url}
\usepackage{subfigure}
\usepackage{verbatim}
\usepackage{natbib}
\usepackage{booktabs}

\usepackage{caption}
\usepackage{makecell}
\usepackage{graphicx}
\usepackage{textcomp}
\usepackage{xcolor}
\usepackage{url}
\usepackage{amsmath,amssymb,amsfonts}
\usepackage{array}
\usepackage{color,framed}
\usepackage{multirow}
\usepackage{ulem}  
\usepackage{setspace}
\usepackage{comment}
\usepackage{pifont}

\newcommand{\reffig}[1]{Fig.\ref{#1}}
\newcommand{\refeqs}[1]{Eq.\ref{#1}}

\def\BibTeX{{\rm B\kern-.05em{\sc i\kern-.025em b}\kern-.08em
    T\kern-.1667em\lower.7ex\hbox{E}\kern-.125emX}}
\usepackage{amsmath}
\usepackage{graphicx}
\usepackage{pgfplots}

\theoremstyle{thmstyleone}%

\theoremstyle{thmstyletwo}%

\theoremstyle{thmstylethree}%

\begin{document}

\title[A Generic Framework for Optimization in Blockchain Simulators]{A Generic Framework for Optimization in Blockchain Simulators}

\author[1]{ \sur{Hou-Wan Long}}\email{houwanlong@link.cuhk.edu.hk}

\author[2]{ \sur{Yujun Pan}}\email{mc36604@um.edu.mo}

\author[2]{ \sur{Xiongfei Zhao}}\email{yb97480@um.edu.mo}

\author*[2]{ \sur{Yain-Whar Si}}\email{fstasp@um.edu.mo}

\affil[1]{\orgdiv{The Chinese University of Hong Kong, Hong Kong}}

\affil*[2]{\orgdiv{University of Macau, Macau}}

\abstract{As blockchain technology rapidly evolves, researchers face a significant challenge due to diverse and non-standardized simulation parameters, which hinder the replicability and comparability of research methodologies. This paper introduces a Generic Framework for Optimization in Blockchain Simulators (GFOBS), a comprehensive and adaptable solution designed to standardize and optimize blockchain simulations. GFOBS provides a flexible platform that supports various optimization algorithms, variables, and objectives, thereby catering to a wide range of blockchain research needs. The paper's key contributions are threefold: the development of GFOBS as a versatile tool for blockchain simulation optimization; the introduction of an innovative optimization method using warm starting technique; and the proposition of a novel concurrent multiprocessing technique for simultaneous simulation processes. These advancements collectively enhance the efficiency, replicability, and standardization of blockchain simulation experiments.}

\keywords{Blockchain, Machine learning, Optimisation, Simulation, Warm start}

\maketitle

\section{Introduction}
As blockchain technology continues to evolve in complexity and scope, the blockchain research community is increasingly confronted with the need for innovative and effective optimization strategies. This demand is widespread across various research areas, including anomaly detection \cite{podgorelec2019machine}, smart contract analysis \cite{liao2019soliaudit}, consensus mechanism optimization \cite{sanghami2022machine}, and cryptocurrency price prediction \cite{chen2020bitcoin}. However, a significant challenge emerges in the realm of replicability and validation of experiments, particularly those employing diverse optimization techniques in blockchain simulations.

Davison's study \cite{Davison2010}, for instance, underscores the complexities in replicating computational experiments, especially in intricate fields like computational neuroscience. Similarly, study by Črepinšek et al. \cite{Crepinsek2014} highlights the specific challenges in replicating and comparing computational experiments. These studies collectively emphasize the need for greater standardization and methodological rigor, resonating strongly with the challenges in blockchain simulation optimization.

In blockchain research, the diverse array of optimization algorithms, variables, and objectives used in different studies further complicates this issue. Each study might adopt unique optimization algorithms, focus on different optimization variables, and pursue distinct optimization objectives. While this diversity enriches the research landscape. Variations in methodologies, simulator settings, and optimization criteria across studies add to the complexity of replicating and validating research findings.

In response to these intricate challenges, we introduce a Generic Framework for Optimization in Blockchain Simulators (GFOBS). This framework is a solution designed to bridge the gap highlighted by these challenges. GFOBS is a generic and inclusive framework that aims streamline the process of blockchain simulation optimization. It offers researchers the flexibility to employ various optimization algorithms and to choose from numerous optimization variables and objectives, catering to the specific requirements of different blockchain optimization problems. GFOBS is also designed to be compatible with Blockchain simulators such as SIMBLOCK \cite{SIMBLOCK}, BlockSim \cite{BlockSim} and Bitcoin-Simulator \cite{gervais2016security}. By providing a standardized approach, GFOBS addresses the critical need for replicability and consistency in blockchain research, ensuring that findings are not only optimized for performance but also consistent and replicable across various studies and scenarios.

The GFOBS encompasses the following key aspects:
\begin{itemize}
    \item Flexibility in Optimization Algorithms: It offers a selection of optimization algorithms, ranging from traditional methods to advanced machine learning techniques.
    \item Variability in Optimization Variables: The framework supports the use of a wide array of optimization variables, including but not limited to block size, transaction processing rules, network latency, mining difficulty, crucial for the performance and efficiency of blockchain networks.
    \item Defining Optimization Objectives: Researchers can precisely define their optimization objectives, whether to maximize efficiency, minimize cost, enhance transaction throughput, improve security, or other goals pertinent to blockchain contexts.
    \item \textcolor{black}{Interoperability: The GFOBS framework is designed with an interface that supports integration with a variety of blockchain simulators, facilitating compatibility across different simulation environments.} This interface is designed to effectively handle diverse parameters and arguments across different simulation tools. These simulators include SIMBLOCK \cite{SIMBLOCK}, BlockSim \cite{BlockSim} and Bitcoin-Simulator \cite{gervais2016security}.
    \item Enhancing Efficiency: GFOBS introduces innovative techniques such as warm-starting and concurrent multiprocessing simulation to enable faster convergence.
\end{itemize}

The contributions of this paper are significant and multifaceted, encapsulating three major advancements in the field of blockchain technology research:
\begin{itemize}
\item[(i)] \textcolor{black}{Introduced the "Generic Framework for Optimization in Blockchain Simulators" (GFOBS), providing a flexible and adaptable platform for blockchain simulation that supports various optimization strategies, parameters, and goals across different blockchain simulators.}
\item[(ii)] \textcolor{black}{Proposed an innovative warm-starting method using pre-computed parameter values and Jaccard similarity, enhancing the speed and efficiency of convergence in blockchain simulations.}
\item[(iii)] \textcolor{black}{Developed a concurrent multiprocessing approach, allowing for parallel simulations in blockchain environments, significantly improving process efficiency.}
\end{itemize}

\textcolor{black}{The remainder of this paper is structured as follows: Section 2 provides a summary of related work. Section 3 details the GFOBS framework. Section 4 explain the optimization process within GFOBS framework. Section 5 presents the experimental setup and outcomes. Section 6 concludes the paper.}

\section{\textcolor{black}{Related work}}
This section reviews existing blockchain simulators, their optimizations, and optimization techniques from other domains to contextualize our research within the current landscape, identify gaps our work addresses, and incorporate cross-domain optimization strategies to enhance the robustness and innovation of our blockchain simulation framework.

\subsection{Existing Blockchain Simulators:}
Blockchain simulation tools are vital for evaluating the performance and security of blockchain networks, enabling researchers to analyze and refine blockchain protocols under various conditions. SIMBLOCK \cite{SIMBLOCK}, developed by Aoki et al., is a notable simulator that facilitates the modification of node behaviors to investigate their impact on the blockchain network's performance. This tool has been instrumental in understanding the effects of neighbor node selection algorithms and relay networks on block propagation time.

BlockSim \cite{BlockSim}, presented by Faria and Correia, is another significant simulator offering a discrete-event environment for modeling and evaluating various blockchain configurations, aiding in the optimization of blockchain performance. Additionally, BlockPerf extends the capabilities of existing simulators like BlockSim, offering a more realistic and detailed analysis of blockchain performance. Polge et al. demonstrated that BlockPerf could provide more accurate results, improving the outcomes by approximately 50\% compared to its predecessors \cite{Polge2021BlockPerf}. Another contribution to the field is Zelig \cite{erdogan2021demo}, a customizable blockchain simulator highlighted by Erdogan et al., which stands out for its ability to simulate custom network topologies without altering the simulator's code, thus offering greater flexibility for research and development in blockchain networks. The systematic mapping study conducted by Albshri et al. offers a comprehensive overview of the blockchain simulation landscape, discussing the quality, capabilities, and features of existing simulators \cite{Albshri2022BlockchainSimulators}. This study underlines the need for high-quality simulators to explore the wide range of features and capabilities of blockchain technologies effectively. Furthermore, the work by Zheng et al. on blockchain simulators for IoT environments provides a focused review on simulators capable of mimicking blockchain networks within IoT settings, presenting a critical analysis of their advantages and limitations \cite{Zheng2022BlockchainSimulatorsIoT}. These simulators and studies form a foundation for ongoing research and development in blockchain technologies, providing essential tools and insights that enable the systematic testing, validation, and optimization of blockchain networks.

Nevertheless, many current blockchain simulators, such as SIMBLOCK and BlockSim, focus on network behavior and transaction processing without deeply integrating optimization algorithms for performance enhancement. GFOBS could fill this gap by offering a more comprehensive and integrated optimization approach that enhances both the simulation efficiency and the accuracy of blockchain network modeling.

% \textcolor{blue}{Blockchain simulation tools play a critical role in evaluating the performance, scalability, and security of blockchain protocols under controlled environments. A variety of simulators have been proposed in the literature, each offering distinct capabilities.}

% \textcolor{blue}{SIMBLOCK \cite{SIMBLOCK}, developed by Aoki et al., enables researchers to simulate block propagation in peer-to-peer networks, emphasizing neighbor node selection and relay strategies. BlockSim \cite{BlockSim} provides a discrete-event simulation environment, facilitating the modeling of consensus protocols, transactions, and network topologies. BlockPerf \cite{Polge2021BlockPerf} extends BlockSim with enhanced realism, offering up to 50\% more accurate results in performance evaluations. Zelig \cite{erdogan2021demo} introduces flexibility through customizable network topologies without modifying the core simulation logic. Additionally, systematic studies like those by Albshri et al. \cite{Albshri2022BlockchainSimulators} and Zheng et al. \cite{Zheng2022BlockchainSimulatorsIoT} provide overviews of blockchain simulators for general and IoT-specific applications, respectively.}

% Nevertheless, many current blockchain simulators, such as SIMBLOCK and BlockSim, focus on network behavior and transaction processing without deeply integrating optimization algorithms for performance enhancement. GFOBS could fill this gap by offering a more comprehensive and integrated optimization approach that enhances both the simulation efficiency and the accuracy of blockchain network modeling.

\subsection{Optimizations in Blockchain Simulators}
Optimization in blockchain simulators predominantly revolves around enhancing the execution of consensus algorithms and improving the efficiency of data storage and processing. The research by Zhang and Xu \cite{Zhang2023OptimizationBlockchain} showcases the application of genetic algorithms for optimizing the supply chain in a blockchain context, highlighting the potential of optimization algorithms to improve the logistics and distribution aspects within blockchain frameworks.

Moreover, the optimization of storage mechanisms in blockchain is addressed by Zeng et al. \cite{Zeng2018StorageOptimization}, where they propose a storage optimization algorithm based on Pearson similarity and K-means clustering to enhance the efficiency of data handling in blockchain applications. This approach not only optimizes the storage but also contributes to the better management of data within the blockchain, signifying the importance of algorithmic optimization in blockchain technology. Further, the work of Feng et al. \cite{Feng2020OptimizationMEC} on joint optimization of radio and computational resources in blockchain-enabled mobile edge computing systems emphasizes the need for a holistic approach to optimization in blockchain, where both computational and network resources are optimized simultaneously to achieve a balanced performance. In the context of blockchain's operational efficiency, the study by Wu \cite{Wu2023OptimizationInterconnection} on the optimization of intelligent interconnection systems using blockchain technology proposes methods to improve system performance through optimized consensus mechanisms and enhanced data throughput. These studies illustrate the multifaceted nature of optimization in blockchain simulators and applications, where different aspects, from supply chain logistics to computational and storage resources, are optimized to enhance the overall performance and reliability of blockchain systems.

However, while studies like those by Zhang and Xu \cite{Zhang2023OptimizationBlockchain} touch upon optimization in blockchain contexts, they often concentrate on specific aspects like supply chain management. GFOBS can bridge this gap by providing a generic optimization framework that is not limited to a particular blockchain application but can be applied broadly across various blockchain simulation scenarios, thereby enhancing the generalizability and utility of blockchain simulators.

\subsection{Optimization in Simulators from Other Domains}
Optimization in simulators extends beyond blockchain, covering a wide array of domains where the enhancement of performance and accuracy is crucial. In the broader field of simulation optimization, techniques are developed to integrate optimization algorithms with simulation models, aiming to boost the system's overall performance. The work of \cite{Amaran2014SimulationOptimization} delves into various algorithms and their applications within simulation optimization, providing insights into the complexities and future prospects of this area.

Further contributions in this domain include the efforts of \cite{Carson1997SimulationOptimization}, who discuss the methods and applications of simulation optimization, underlining its importance in efficiently conducting simulation experiments to derive maximum information with minimal resource expenditure. Similarly, \cite{Schruben1986FrequencyDomain} introduces frequency-domain methods in simulation experiments, offering a novel approach to optimize simulation processes and enhance decision-making based on simulation results. In the field of transportation, \cite{Osorio2017SimulationBasedOptimization} discuss the use of multiple simulators to address urban transportation problems, showcasing how combining various simulation models can lead to optimized solutions for complex urban transportation systems. Likewise, \cite{Petrasinovic2018GeometryOptimization} applied genetic algorithms for the geometry optimization of flight simulator mechanisms, demonstrating how optimization can significantly impact the design and functionality of simulators in aeronautical training. Additionally, simulation optimization is pivotal in manufacturing and industrial applications, as demonstrated by \cite{Hani2008SimulationOptimizationTrain}, who utilized simulation-based optimization methods to enhance the scheduling policies in a railway maintenance facility. This approach exemplifies how optimization can be employed to improve operational efficiencies and decision-making in industrial settings. These studies and developments across different domains underline the significance of optimization in simulation, showcasing its ability to enhance the fidelity, efficiency, and accuracy of simulations. As these optimization techniques continue to evolve, they contribute profoundly to the advancement of simulation technologies, enabling more accurate predictions and better performance across various fields.

However, optimization in non-blockchain simulators often involves advanced algorithms and techniques that are not yet standard in blockchain simulation. By incorporating such sophisticated optimization methods, GFOBS can fill the gap by bringing a new level of efficiency and effectiveness to blockchain simulations, drawing on the best practices and innovations from a wider range of simulation disciplines.

Next section will shown the various facets and operational dynamics of the GFOBS. For the purpose of simplification, the notations used in this paper are summarized in Table~\ref{notations}.

\vspace{-0.3cm}
\linespread{1}
\begin{table}
    \centering
    \scriptsize
    \caption{\textcolor{black}{Notations used in this paper}}
%   \resizebox{0.96\hsize}{!}{
     %\begin{threeparttable}
    %\resizebox{0.96\hsize}{!}{
         \begin{tabular}{p{1cm}<{\raggedright} | p{11cm}<{\raggedright}}
            \hline\hline
%           \specialrule{0em}{3pt}{1pt}
%           \textcolor{black}{\text{Notation}} & \textcolor{black}{\text{Definition}} \\
%           \specialrule{0em}{3pt}{1pt}
%           \hline  
%           \specialrule{0em}{1pt}{1pt} 
%           \textcolor{black}{$T$} & \textcolor{black}{Throughtput} \\
%           \specialrule{0em}{1pt}{1pt}
%           \hline
%           \specialrule{0em}{1pt}{1pt} 
%           \textcolor{black}{$\sigma$}  & \textcolor{black}{Block incentive volatility}\\
%           \specialrule{0em}{1pt}{1pt}
%           \hline
            \specialrule{0em}{1pt}{1pt} 
            \specialrule{0em}{1pt}{1pt} 
            \textcolor{black}{$\omega_j$} & \textcolor{black}{Optimisation objective} \\
            \specialrule{0em}{1pt}{1pt}
            \hline
            \specialrule{0em}{1pt}{1pt} 
            \textcolor{black}{$s_{\omega_j}$} & \textcolor{black}{Weight of optimisation objective} \\
            \specialrule{0em}{1pt}{1pt}
            \hline
            \specialrule{0em}{1pt}{1pt} 
            \textcolor{black}{$\Omega$} & \textcolor{black}{The set of optimisation objective} \\
            \specialrule{0em}{1pt}{1pt}
            \hline
            \specialrule{0em}{1pt}{1pt} 
            \textcolor{black}{$\phi_i$} & \textcolor{black}{Variable parameter} \\
            \specialrule{0em}{1pt}{1pt}
            \hline
            \specialrule{0em}{1pt}{1pt} 
            \textcolor{black}{$R_{\phi_i}$} & \textcolor{black}{The Range of Variable parameter} \\
            \specialrule{0em}{1pt}{1pt}
            \hline
            \specialrule{0em}{1pt}{1pt} 
            \textcolor{black}{$\Phi$} & \textcolor{black}{The set of variable parameters} \\
            \specialrule{0em}{1pt}{1pt}
            \hline
            \specialrule{0em}{1pt}{1pt} 
            \textcolor{black}{$f_{\phi_i}$} & \textcolor{black}{The value of fixed parameter $\phi_i$} \\
            \specialrule{0em}{1pt}{1pt}
            \hline
            \specialrule{0em}{1pt}{1pt} 
            \textcolor{black}{$\psi_k$} & \textcolor{black}{Warm-starting parameters configuration} \\
            \specialrule{0em}{1pt}{1pt}
            \hline
            \specialrule{0em}{1pt}{1pt} 
            \textcolor{black}{$\Psi$} & \textcolor{black}{The set of warm-starting parameters configuration} \\
            \specialrule{0em}{1pt}{1pt}
            \hline
            \specialrule{0em}{1pt}{1pt} 
            \textcolor{black}{$c_{\phi_i}^{\psi_k}$} & \textcolor{black}{The value of parameter $\phi_i$ in warm-starting parameters configuration $\psi_k$} \\
            \specialrule{0em}{1pt}{1pt}
            \hline
            \specialrule{0em}{1pt}{1pt} 
            \textcolor{black}{$R_{\phi_i}^{\psi_k}$} & \textcolor{black}{The range of parameter $\phi_i$ in warm-starting parameters configuration $\psi_k$} \\
            \specialrule{0em}{1pt}{1pt}
            \hline
            \specialrule{0em}{1pt}{1pt} 
            \textcolor{black}{$\Psi^*$} & \textcolor{black}{The optimal warm-starting parameters configuration set} \\
            \specialrule{0em}{1pt}{1pt}
            \hline
            \specialrule{0em}{1pt}{1pt} 
            \textcolor{black}{$\psi_k^*$} & \textcolor{black}{A configuration in the optimal warm-starting parameters configuration set} \\
            \specialrule{0em}{1pt}{1pt}
            \hline
            \specialrule{0em}{1pt}{1pt} 
            \textcolor{black}{$c_{\phi_i}^{\psi_k^*}$} & \textcolor{black}{The value of parameter $\phi_i$ in optimal warm-starting parameters configuration $\psi^*_k$} \\
            \specialrule{0em}{1pt}{1pt}
            \hline
            \specialrule{0em}{1pt}{1pt} 
            \textcolor{black}{$JS_{\psi_k}$} & \textcolor{black}{The Jaccard similarity of warm-starting parameters configuration $\psi_k$} \\
            \specialrule{0em}{1pt}{1pt}
            \hline
            \specialrule{0em}{1pt}{1pt} 
            \textcolor{black}{$N$} & \textcolor{black}{The population of optimisation algorithm} \\
            \specialrule{0em}{1pt}{1pt}
            \hline
            \specialrule{0em}{1pt}{1pt} 
            \textcolor{black}{$Arg_{\phi_i}$} & \textcolor{black}{The argument of parameter $\phi_i$ passed to simulator} \\
            \specialrule{0em}{1pt}{1pt}
            \hline
            \specialrule{0em}{1pt}{1pt} 
            \textcolor{black}{$p_{\phi_i}^{\psi_k^*}$} & \textcolor{black}{The value of parameter $\phi_i$ in the k-th candidate solution} \\
            \specialrule{0em}{1pt}{1pt}
            \hline
            \specialrule{0em}{1pt}{1pt} 
            \textcolor{black}{$v_{\omega_j}^{\psi_k^*}$} & \textcolor{black}{The result of objective $\omega_j$ in a simulation using the k-th configuration in $\Psi^*$} \\
            \specialrule{0em}{1pt}{1pt}
            \hline
            \specialrule{0em}{1pt}{1pt} 
            \textcolor{black}{$\mathbb{1}_{(\cdot)}^{(\bullet)}$} & \textcolor{black}{Indicator} \\
            \specialrule{0em}{1pt}{1pt}
            \hline
        \end{tabular}
    %}
      %\end{threeparttable}
%   }
    \label{notations}
\end{table}

\section{A Generic Framework for Optimization in Blockchain Simulators}\label{GFOBS}
\begin{figure*}[htp]
    \centering
    \includegraphics[width=1\textwidth]{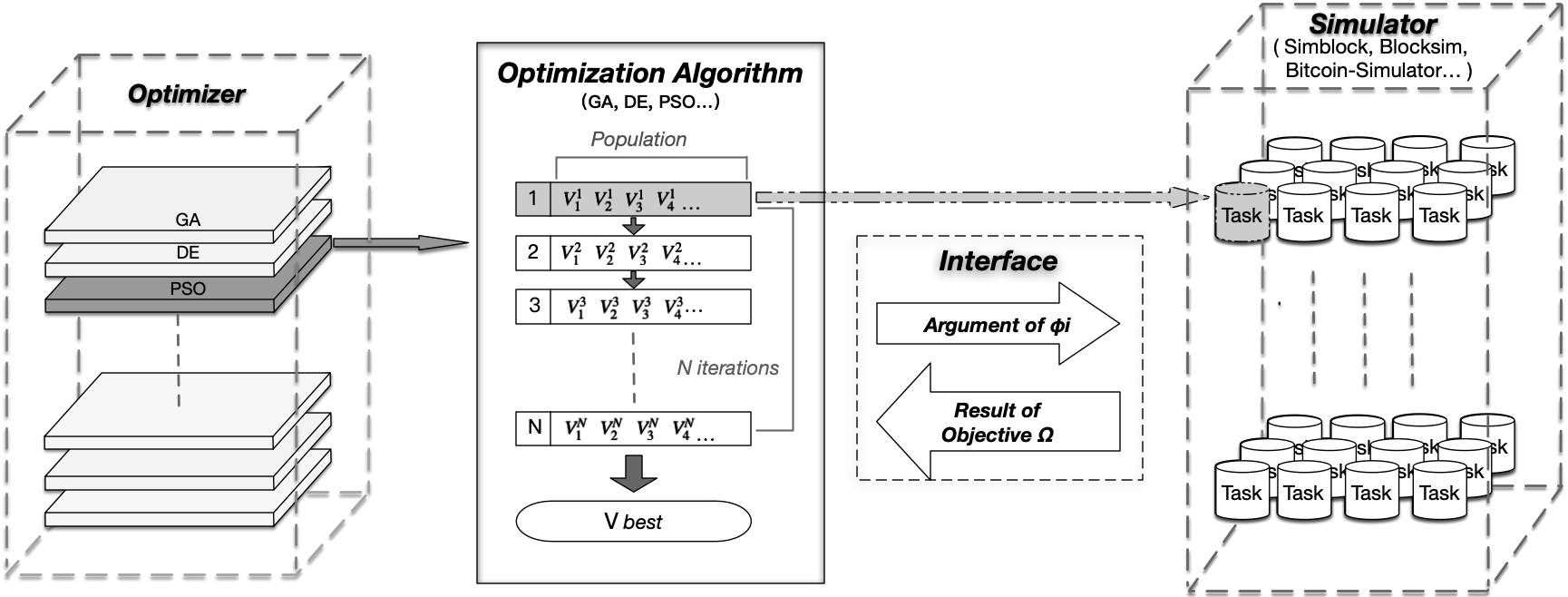}
    \caption{Three components of GFOBS}
    \label{flow}
\end{figure*}

%\subsection{Key Components}
GFOBS comprises three integral components as shown in \reffig{flow}: Interface, Simulator, and Optimizer, each playing a distinct yet interconnected role to enhance blockchain simulations and optimizations. 
GFOBS begins with the user defining key elements: the optimization parameters selected from the set of variable parameters, the specific optimization objective, and the chosen optimization algorithms. Once the optimization process is initiated, the parameter values are conveyed to the blockchain simulator as shown in pseudo-code \ref{GFOBS process}. This transmission is facilitated through a specially designed interface that connects the optimizer and the simulator. The simulator then executes simulations using these parameters. After each simulation run, the outcomes are relayed back to the optimizer for evaluation. This feedback loop allows for continuous assessment and adjustment of parameters, ensuring the optimization process is dynamic and responsive to the simulation results. 
\begin{algorithm}
%\textsl{}\setstretch{1.8}
\begin{scriptsize}
\floatname{algorithm}{Pseudo-code}
%\begin{algorithm} [H]
\caption{GFOBS}
\label{GFOBS process}
\begin{algorithmic}[1]
\Procedure{Optimisation on GFOBS}{}
    \Procedure{The warm-starting procedure}{}
        \State Return $\Psi^*$
    \EndProcedure
    \Procedure{Optimisation}{}
        \State Generate $\{Args^{\psi_k^*}_{init}|\psi_k^*\in\Psi^*\}$ by \refeqs{init}
        \State Pass $\{Args^{\psi_k^*}_{init}|\psi_k^*\in\Psi^*\}$ to SIMBLOCK
        \Procedure{Simulation}{}
            \State Return $\{\{v_{\omega_j}^{\psi_k^*}|\omega_j\in\Omega\}|\psi_k^*\in\Psi^*\}$
        \EndProcedure
        \State Calculate $U^{\psi_k^*}(\Omega)$ for each element in $\{\{v_{\omega_j}^{\psi_k^*}|\omega_j\in\Omega\}|\psi_k^*\in\Psi^*\}$
        \State Generate new sets of value for optimisation parameters $\{\{p_{\phi_1}^{\psi_k^*},p_{\phi_2}^{\psi_k^*},\dots,p_{\phi_{|\Phi|}}^{\psi_k^*}\}|\psi_k^*\in\Psi^*\}$
        \State Generate $\{Args^{\psi_k^*}_{subs}|\psi_k^*\in\Psi^*\}$ by \refeqs{subs}
        \State Pass $\{Args^{\psi_k^*}_{subs}|\psi_k^*\in\Psi^*\}$ to SIMBLOCK
        \State Repeat step 8-14 until a stopping criteria is met
    \EndProcedure
\EndProcedure
\end{algorithmic}
%\end{algorithm}
\end{scriptsize}
\end{algorithm}

\subsection{Simulator}
A blockchain simulator emulates the behavior of a real blockchain network. This includes the simulation of nodes in the network, their interactions, and the transmission of data. By replicating the dynamics of a blockchain, the simulator provides a controlled environment for analysis and experimentation. As an integral component of the proposed GFOBS, blockchain simulators facilitate the testing of parameters, algorithms, and protocols. We have made cross-comparison on SIMBLOCK \cite{SIMBLOCK}, BlockSim \cite{BlockSim} and Bitcoin-Simulator \cite{gervais2016security} and filter for those common parameters, which are summarized in Table \ref{blockchain simulator parameters}. In the experiment section, these parameters will be unified before the experiment to minimize the effects caused by different simulators. Parameters that are unique to a specific simulator will retain their default status.
\begin{table*}[t]
\scriptsize

    \centering
    \caption{Common parameters across \textcolor{black}{SIMBLOCK \cite{SIMBLOCK}, BlockSim \cite{BlockSim} and Bitcoin-Simulator \cite{gervais2016security}}}
    \label{blockchain simulator parameters}
    \begin{threeparttable}

    \renewcommand{\arraystretch}{1.5} % Increase line spacing
    \begin{tabular}{m{2cm}<{\raggedright} | m{4.5cm}<{\raggedright} | m{5.5cm}<{\raggedright}}
    \toprule
    Parameter & Description & Initial Value \\
    \midrule
     \texttt{BLOCK HEIGHT}& the number of blocks generated & 100\\
    \hline 
    \texttt{BLOCK SIZE}& the size of blocks generated (unit: megabyte) & 1\\
    \hline 
    \texttt{EXPECTED MINING INTERVAL}& the average time taken to mine a new block (unit: second) & 600\\
    \hline 
    \texttt{AVERAGE HASH RATE}& the average number of hash calculation executed per millisecond of each node (unit: 1/millisecond) & 40,000\\
    \hline 
    \texttt{NODE NUMBER}& the number of nodes participating in the blockchain network & 6,000\\
    \hline 
    \texttt{REGION DISTRIBUTION}\tnote{a}& the distribution of node's region. Each value means the rate of the number of nodes in the corresponding region to the number of all nodes & $
        {\left\{
        \begin{array}{cccccc}
  0.3869 & 0.5159 & 0.0113 &\\
  0.0574 & 0.0119 & 0.0166
\end{array}
        \right\}
        }$\\
    \hline 
    \texttt{LATENCY}& the list of latency from one region i to region j (unit: millisecond) & \makecell[l]{ $
        {\left\{
        \begin{array}{cccccc}
  32 & 124 & 184 & 198 & 151 & 189 \\
  124 & 11 & 227 & 237 & 252 & 294 \\
  184 & 227 & 88 & 325 & 301 & 322 \\
  198 & 237 & 325 & 85 & 58 & 198 \\
  151 & 252 & 301 & 58 & 12 & 126 \\
  189 & 294 & 322 & 198 & 126 & 16
\end{array}
        \right\}
        }
    $}\\
    \hline 
    \texttt{UPLOAD BANDWIDTH}\tnote{b}& the list of upload bandwidth assigned to each region (unit: bit per second) & $\begin{array}{ccccccc}
  \{19200000 & 20700000 & 5800000 & 15700000\\ & 10200000 & 11300000 & 6*100000\}
\end{array}$ \\
    \hline 
    \texttt{DOWNLOAD BANDWIDTH}\tnote{c}& the list of download bandwidth assigned to each region (unit: bit per second) &  $\begin{array}{ccccccc}
  \{52000000 & 40000000 & 18000000 & 22800000\\ & 22800000 & 29900000 & 6*100000\}
\end{array}$ \\
    \bottomrule
    \end{tabular}
    \begin{tablenotes}
\scriptsize
\item[a] regions considered are "NORTH AMERICA", "EUROPE", "SOUTH AMERICA", "ASIA PACIFIC", "JAPAN", "AUSTRALIA" respectfully
\item[b] last element is inter-regional upload bandwidth
\item[c] last element is inter-regional download bandwidth
\end{tablenotes}
    \end{threeparttable}
\end{table*}

The Simulator within GFOBS is crucial for emulating blockchain processes like block creation. It interacts with the optimizer, adjusting its block generation in response to the optimizer's parameters. This capability is fundamental for conducting precise and realistic blockchain simulations, thereby playing a significant role in GFOBS. The design of GFOBS accommodates the integration of multiple simulators, underscoring its versatility and comprehensive application in blockchain studies.

\subsection{Interface}
The Interface in GFOBS acts as an essential mediator between the optimizer and the blockchain simulator. It is engineered for compatibility with a wide array of simulators and optimizers, regardless of their unique configurations or algorithms. This component efficiently facilitates the exchange of optimization parameters as arguments and simulation results, ensuring seamless communication and adaptability across various blockchain simulation platforms.

In the interface, Python's subprocess module is utilized to facilitate the interaction between the optimiser and the simulator. This module manages the execution of the simulator and the retrieval of its output, essential for the optimization process. The subprocess module employs \texttt{subprocess.run()}, a function that initiates the simulator with parameters from the optimiser and captures the resulting output. The interaction is streamlined as follows:
\begin{enumerate}
\item[(i)] \text{Parameter Passing}: Optimized parameters are passed as arguments to the simulator's executable via \texttt{subprocess.run()}.
\item[(ii)] \text{Execution and Response}: The simulator processes these parameters, executing the blockchain simulation and generating output data.
\item[(iii)] \text{Data Retrieval and Analysis}: The module captures this output, which the optimiser then analyzes to inform further optimization steps.
\end{enumerate}

This use of the subprocess module in GFOBS's interface component illustrates an effective integration of simulator and optimiser, enhancing the framework’s overall efficiency and scalability.

\subsection{Optimizer}

%\subsection{Optimisation algorithms}\label{warmstart}
The primary goal in developing GFOBS is to enhance research flexibility by allowing the use of various optimization algorithms, the selection from a wide array of optimization variables, and the definition of diverse optimization objectives. This flexibility is essential for addressing the specific and multifaceted nature of blockchain-related optimization challenges.
For a clearer understanding of GFOBS's optimization process, we define the following terms:
\begin{itemize}
    \item[(i)] \textit{\text{Optimisation objective:} The specific goal set by the researcher, such as maximizing throughput or minimizing latency in a blockchain network. It is denoted by $\omega_j$. The set of optimisation objective is denoted by $\Omega$}
    
    \item[(ii)] \textit{\text{Variable parameters:} These are parameters within the GFOBS that can be varied during the optimization process, \textcolor{black}{such as block size, expected mining interval in a blockchain simulator.} The set of all optimizable parameters is represented by $\Phi$.}
    
    \item[(iii)] \textit{\text{Optimisation parameters:} These are a subset of the \textcolor{black}{Variable} parameters that are actively varied in the optimization process to achieve the stated objective. For instance, if maximizing throughput is the goal, and the chosen parameters are block size and expected mining interval, then these become the optimization parameters. The notation $\mathbb{1}_{\phi_i}^\Phi=1$ indicates that the optimizable parameter $\phi_i$ is being used as an optimization parameter.}
    
    \item[(iv)] \textit{\text{Fixed parameters:} Parameters within GFOBS that are potential variables for optimization but are assigned fixed values by the user. For example, in optimizing throughput with block size and expected mining interval as variables, other parameters would be set to fixed values. For a fixed parameter $\phi_i$, $\mathbb{1}_{\phi_i}^\Phi=0$, and its value is denoted by $f_{\phi_i}$. The indicator function for a parameter $\phi_i$ is defined as:}
    \begin{equation}
        \mathbb{1}_{\phi_i}^\Phi=\left\{
        \begin{aligned}
            &1, \quad \textit{if $\phi_i$ is optimisation parameter} \\
            &0, \quad \textit{if $\phi_i$ is fixed parameter}
        \end{aligned}\right.
    \end{equation}
\end{itemize}

The Optimizer in GFOBS encompasses several optimization algorithms, optimization objectives and optimisation parameters. It operates by evaluating the optimization objective through a predefined objective function, subsequently generating refined optimization parameter configurations. This component is vital for the iterative enhancement of blockchain simulations, aligning them closely with specified research objectives and ensuring optimal performance outcomes in blockchain explorations.

The optimization algorithms that have been tested in GFOBS are Genetic Algorithm (GA), Differential Evolution (DE), and Particle Swarm Optimization (PSO), which are all evolutionary algorithms and suitable for the multi-dimensional aspect of blockchain simulator parameter optimization.
As shown in Table \ref{tab:optimization_algorithms}, in these three algorithms, parameters such as the maximum number of iterations, population size, and the lower and upper bounds (lb and ub) are common. 

The termination condition is defined by a fixed maximum number of iterations, depending on the complexity of the objective function. Larger population size tends to promote exploration, allowing the algorithm to search a wider range of solutions, smaller population size can lead to faster convergence, but also risk missing a wider optimal region due to under exploration. The lb and ub parameters represent the minimum and maximum boundaries of the population, essentially defining the range within which the optimal solution for the parameters is sought.

\textcolor{black}{ For the Genetic Algorithm, there are other two basic and key parameters: Crossover rate and Mutation rate. Crossover rate, ranging from 0 to 1, dictates the frequency of crossover events per generation in a genetic algorithm, where 1 means every offspring results from crossover and 0 means no crossover occurs, with offspring directly copied from the previous generation except for mutations.  Mutation rate is in the range of [0, 1], it plays a crucial role in maintaining genetic diversity within a population. It ensures that the algorithm explores a wider search space and avoids premature convergence to a local optimum \cite{hassanat2019choosing}. }

\textcolor{black}{ For the Differential evolution (DE) algorithm, according to \cite{mallipeddi2011differential}, it leverages a set of potential solution vectors, each with \( D \) dimensions, known as individuals in a population size \( NP \). The initial vectors are distributed across the search space within given bounds \( X_{\text{min}} \) and \( X_{\text{max}} \). An individual's parameter at generation \( G=0 \) is computed as \refeqs{de},
\begin{equation}
\label{de}
x_{i,0}^j = x_{\text{min}}^j + \text{rand}(0, 1) \times (x_{\text{max}}^j - x_{\text{min}}^j), \quad j = 1, \ldots, D
\end{equation}
where \(\text{rand}(0, 1)\) generates a uniformly distributed number between 0 and 1. This ensures a diverse starting point for the optimization process. Besides, DE algorithm's main parameters include the scale factor for mutation($F$) and Crossover Rate($CR$). The scaling factor $F$ is a positive control parameter for scaling the difference vectors. It affects the creation of donor vectors and the direction of searches within the population.} 

The PSO algorithm can be shown using \refeqs{pso1} and \refeqs{pso2}, k denote the iteration number, $\omega$ represent the inertia weight factor, and c1 as well as c2 stand for constants representing the control parameters. Additionally, r1 and r2 denote random values uniformly distributed in the range [0, 1] \cite{mahdi2010parameter}. $\omega$ adjusts the momentum of the particle, higher w promotes exploration. $c_1$ improves the solution by observing the particle's own best position. An increased $c_1$ value steers the optimization process more strongly toward finding individual optimal solutions. $c_2$ uses the group's best findings to guide the search. A higher $c_2$ emphasizes the role of shared knowledge between particles, pushing optimization toward the best solution found by the group \cite{wang2018particle}.

\begin{equation}
\label{pso1}
\begin{split}
v_i(k + 1) = & \omega v_i(k) + c_1 \text{r}_1(p_i(k) - x_i(k)) \\
& + c_2 \text{r}_2(p_g(k) - x_i(k))
\end{split}
\end{equation}

\begin{equation}
\label{pso2}
x_i(k + 1) = x_i(k) + v_i(k + 1)
\end{equation}

\begin{table}[t]
\scriptsize
    \centering
    \caption{Parameters for Optimization Algorithms}
    \label{tab:optimization_algorithms}
    \begin{tabular}{c|c|c|c}
    \toprule
    \textbf{ Algorithm } & \textbf{ Parameter } & \textbf{ Value } & \textbf{ Description }\\
    \midrule
       & size\_pop & 50 & Population Size\\
       & max\_iter & 200 & Max number of Iterations\\
    GA & prob\_mut & 0.001 & Mutation Probability\\
       & lb & [1s, 1KB] & Lower bound of Variables\\
       & ub & [1800s, 50MB] & Upper bound of Variables\\
    \hline
       & $F$ & 0.5 & Scale factor for Mutation \\
       & size\_pop & 50 & Population Size\\
    DE & max\_iter & 200 & Max number of Iterations \\
       & $CR$ & 0.3 & Crossover Rate \\
       & lb & [1s, 1KB] & Lower bound of Variables\\
       & ub & [1800s, 50MB] & Upper bound of Variables\\
    \hline
        & pop & 50 & Population Size \\
        & max\_iter & 150 & Max number of Iterations\\
        & $\omega$& 0.5 & Inertia Weight  \\
    PSO & c1 & 0.8 & Personal best Weight \\
        & c2 & 0.8 & Global best Weight\\
        & lb & [1s, 1KB] & Lower bound of Variables  \\
        & ub & [1800s, 50MB] & Upper bound of Variables \\
    \bottomrule
    \end{tabular}
\end{table}

\section{\textcolor{black}{Optimization process}}
Within GFOBS, both the simulator and the optimizer are designed to be replaceable with alternative options. This means users can choose different simulators and optimizers, each with its own set of optimization algorithms, objectives, and parameters. However, it's important to make careful choices regarding these components. Selecting the most suitable simulator and optimizer is key to enhancing the efficiency of both the simulation and optimization processes within GFOBS.

\subsection{Optimization Parameter Selection}

In GFOBS, selecting optimization parameters is a crucial step that aligns the simulation with the user’s specific research objectives. This process enables targeted exploration of blockchain behaviors under varying conditions. For example, a researcher studying the impact of transaction volume on network stability may select parameters such as transaction rate and block verification time. Alternatively, to analyze performance under load, one might choose network load and transaction processing capacity. These parameters become the focus of optimization, allowing direct assessment of their influence on system behavior.

Once variable parameters are selected, fixed values must be assigned to other system aspects—such as average hash power or node count—to ensure a controlled and consistent simulation environment. This selective approach enhances research depth by isolating specific factors and observing their effects. For instance, adjusting block size to evaluate changes in throughput can help identify optimal configurations for various network conditions. Overall, GFOBS empowers researchers to systematically uncover relationships between blockchain parameters, supporting more efficient and insightful simulations.

\subsection{Optimisation objective selection}
In GFOBS, the Optimization Objective Selection feature provides users with extensive flexibility to define their research goals, allowing for a nuanced approach to blockchain optimization. This feature accommodates a broad spectrum of objectives, from straightforward single-goal optimizations to more complex multi-objective strategies, catering to the diverse requirements of blockchain research. For instance, a researcher focused on network security might target the reduction of unauthorized transactions as a singular objective, analyzing how various security protocols affect the incidence of such transactions within the blockchain network.

Expanding on the multi-objective optimization, consider a user whose research entails optimizing both the latency and resilience of a blockchain system. In this context, the user could simultaneously target a decrease in block propagation time and an increase in the system's ability to withstand high transaction loads. This dual-objective approach enables a comprehensive analysis of how these two factors interplay and affect the overall robustness and efficiency of the blockchain.

Moreover, GFOBS allows for dynamic adjustment of these objectives as the research evolves. For example, after initial simulations, a user might discover that reducing block size significantly improves transaction speed. This insight could lead to a refined optimization goal, focusing on determining the smallest block size that maintains network integrity without compromising performance.

The optimization objectives in GFOBS are not just theoretical targets; they are integral to shaping the simulation's direction and focus. By carefully selecting and adjusting these objectives, researchers can ensure that their investigations are aligned with their specific research questions, yielding data and insights that are directly applicable to solving real-world blockchain challenges. For example, in developing a blockchain solution for supply chain management, a researcher could prioritize traceability and transaction verifiability to ensure the authenticity of goods through the supply chain.

\subsection{Optimization Algorithm Selection}

In GFOBS, selecting an appropriate optimization algorithm is critical to accurately addressing diverse blockchain research objectives. Given the complexity of blockchain systems, the algorithm must align with the specific demands of the simulated environment.

For instance, Particle Swarm Optimization (PSO) is well-suited for tasks requiring rapid convergence and fine-tuning in multi-dimensional spaces—such as optimizing smart contract execution without compromising throughput. PSO helps identify the optimal trade-off between contract complexity and network performance. Differential Evolution (DE) excels in exploring complex, high-dimensional spaces and is ideal for balancing transaction speed with network robustness. By systematically navigating the solution space, DE aids in identifying configurations that optimize throughput while preserving security. Genetic Algorithms (GA) are effective for multi-objective scenarios, such as reducing latency while maintaining data integrity. Leveraging evolutionary principles, GA iteratively evolves solutions, making it suitable for refining consensus protocols or enhancing operational efficiency.

Users must also define the population size \( N \), which significantly impacts the search process. A larger \( N \) allows broader exploration, increasing the chance of identifying diverse, high-quality solutions. For example, in scalability-focused tasks, a larger population aids in uncovering resilient network configurations that support efficient scaling.

\subsection{\textcolor{black}{Initialization of simulation}}
The initialization of simulations in our framework is designed to ensure an efficient starting point, leveraging the optimal warm-starting configurations denoted as $\psi_k^*$. These configurations form the basis for generating the initial argument sets $Args^{\psi_k^*}_{\text{init}}$, which are crucial for the first phase of the simulation process. The configuration set $\psi_k^*$, detailed in Section \ref{warm_start}, consists of parameter configurations that have been identified as optimal in previous simulations, providing a proven starting ground for new simulation iterations.

\begin{equation}
\label{init}
\begin{aligned}
    &Args_{init}=\{Arg_{\phi_i}|\phi_i\in\Phi\},\\
    &\textit{where }Arg_{\phi_i}=\mathbb{1}_{\phi_i}^\Phi c_{\phi_i}^{\psi_k^*}+(1-\mathbb{1}_{\phi_i}^\Phi)f_{\phi_i}
\end{aligned}
\end{equation}

In \refeqs{init}, $Arg_{\phi_i}$ represents the argument for each parameter $\phi_i$, calculated as a combination of the optimal parameter value $c_{\phi_i}^{\psi_k^*}$ and a fixed parameter value $f_{\phi_i}$, modulated by the presence of $\phi_i$ in the parameter set $\Phi$. This method ensures that the initial simulation state is aligned with previously successful configurations, enhancing the likelihood of obtaining meaningful and optimized results from the outset.

For subsequent simulation rounds, the argument set $Args_{\text{subs}}$ is dynamically updated based on the outcomes of the current simulation cycle, as shown in the following equation:
\begin{equation}
\label{subs}
\begin{aligned}
    &Args_{subs}=\{Arg_{\phi_i}|\phi_i\in\Phi\},\\
    &\textit{where }Arg_{\phi_i}=\mathbb{1}_{\phi_i}^\Phi p_{\phi_i}^{\psi_k^*}+(1-\mathbb{1}_{\phi_i}^\Phi)f_{\phi_i}
\end{aligned}
\end{equation}

In \refeqs{subs}, $p_{\phi_i}^{\psi_k^*}$ represents the optimized value of the parameter $\phi_i$ that the optimizer suggests for the next iteration. This approach allows for iterative refinement of the simulation parameters, ensuring that each simulation cycle builds upon the knowledge and insights gained from the previous ones.

Throughout the simulation process, the system computes the values $v_{\omega_j}$ for each performance metric $\omega_j$ within the set $\Omega$. These metrics are essential for the optimizer’s evaluation, as they provide quantitative feedback on the simulation's performance under the current parameter settings. The outcomes of each simulation are systematically documented in a local file, creating a comprehensive record of the simulation's performance over time. This data repository becomes a valuable asset for ongoing analysis, allowing for retrospective examination of the simulation results and aiding in the refinement of the simulation model and optimization strategy.

\subsection{\textcolor{black}{Warm-starting (WS)}}\label{warm_start}

Warm-starting enhances iterative optimization by leveraging solutions from prior tasks. In GFOBS, despite varying parameters and objectives across problems, previous optima often remain relevant, especially for related optimization tasks. This strategy, exemplified in Bayesian optimization by Poloczek et al. \cite{warm_baye}, accelerates convergence, particularly in stochastic simulations, thus reducing computational cost.

In SIMBLOCK-based blockchain simulations, randomness arises mainly from mining intervals, reflecting block discovery uncertainty. Unlike Poloczek’s work \cite{warm_baye}, where randomness is embedded in optimization variables, SIMBLOCK's randomness emulates blockchain behavior and is not directly optimized. Erhan et al. \cite{pmlr-v9-erhan10a} showed that warm-starting improves gradient-based optimization by offering better initial points. GFOBS extends this to heuristics, enhancing performance amid the complex parameter space of blockchain simulations (e.g., network size, transaction rate, consensus settings).

\textcolor{black}{Effective warm-starting in GFOBS requires two conditions: (i) the current and prior optimization objectives must align to ensure parameter consistency, and (ii) the parameter ranges of stored solutions must fully encompass those of the current task. As shown in Figure \ref{fig:ranges1}, full coverage aids convergence. Partial overlap, as illustrated in Figure \ref{fig:ranges2} where \( A = [a, a'] \) and \( A_n = [a_n, a_n'] \), risks omitting optimal regions (e.g., \( B = [a_n', a'] \)), potentially preventing global optima.}

\pgfplotsset{compat=1.17}
\usetikzlibrary{decorations.pathreplacing}

\begin{figure}[ht]
\centering
\begin{tikzpicture}
    % Draw axes
    \draw[-stealth, thick] (0,0) -- (9,0);

    % Mark the points on the axis
    \foreach \x/\xtext in {1/a_n, 2/a, 6/a', 8/a_n'}
    {
        \draw (\x,3pt) -- (\x,-3pt);
        \node at (\x,-0.5) {$\xtext$};
    }
    
    % Draw braces and labels
    \draw [decorate,decoration={brace,amplitude=5pt},yshift=2pt]
    (1,0) -- (8,0) node [black,midway,yshift=15pt] {$A_n$};
    \draw [decorate,decoration={brace,amplitude=5pt,mirror},yshift=-2pt]
    (2,0) -- (6,0) node [black,midway,yshift=-15pt] {$A$};

    % Highlight the overlapping region
    \draw[ultra thick] (2,0) -- (6,0);

\end{tikzpicture}
\caption{\textcolor{black}{Complete Parameter Range Inclusion for Warm-Starting Optimization}}
\label{fig:ranges1}
\end{figure}
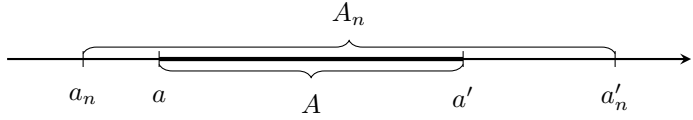

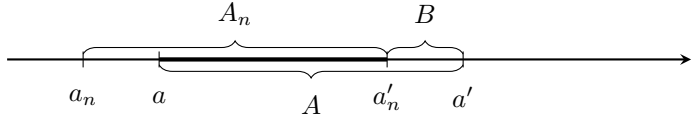
\begin{figure}[ht]
\centering
\begin{tikzpicture}
    % Draw axes
    \draw[-stealth, thick] (0,0) -- (9,0);

    % Mark the points on the axis
    \foreach \x/\xtext in {1/a_n, 2/a, 6/a', 5/a_n'}
    {
        \draw (\x,3pt) -- (\x,-3pt);
        \node at (\x,-0.5) {$\xtext$};
    }
    
    % Draw braces and labels
    \draw [decorate,decoration={brace,amplitude=5pt},yshift=2pt]
    (1,0) -- (5,0) node [black,midway,yshift=15pt] {$A_n$};
    \draw [decorate,decoration={brace,amplitude=5pt,mirror},yshift=-2pt]
    (2,0) -- (6,0) node [black,midway,yshift=-15pt] {$A$};

    \draw [decorate,decoration={brace,amplitude=5pt},yshift=2pt]
    (5,0) -- (6,0) node [black,midway,yshift=15pt] {$B$};

    % Highlight the overlapping region
    \draw[ultra thick] (2,0) -- (5,0);

\end{tikzpicture}
\caption{\textcolor{black}{Partial Parameter Range Inclusion Highlighting Potential Optimization Gaps}}
\label{fig:ranges2}
\end{figure}

The first condition is satisfied by using a warm-starting mechanism that retrieves parameter values from a solution database based on matching optimization objectives. This database, encompassing diverse blockchain cases, ensures broad coverage of potential objectives.

To meet the second condition, the mechanism identifies the top \( N \) populations from past solutions whose parameter ranges best align with the current task. These are selected as the initial population. We quantify this alignment using the Jaccard index, which measures similarity between parameter ranges. Let \( A \) be a parameter range for the current task and \( A_n \) the corresponding range from a prior task. The Jaccard index between them provides a precise similarity metric—higher values indicate better alignment—guiding the selection of the most relevant prior solutions.

\begin{equation}
    \label{Jaccard}
    \text{Jaccard Similarity}(\mathbf{A}, \mathbf{A_n}) = \frac{| \mathbf{A} \cap \mathbf{A_n} |}{| \mathbf{A} \cup \mathbf{{A_n}} |}
\end{equation}

The optimized values in GFOBS are obtained through heuristic algorithms such as Genetic Algorithms (GA) and Differential Evolution (DE), which iteratively refine solutions to improve the objective function—e.g., minimizing transaction validation time or maximizing network throughput. Multiple combinations of optimization objectives and parameter ranges are explored over several iterations, and the resulting optimal values are stored in a database for reuse in future tasks. To better illustrate the warm-starting process in GFOBS, we introduce the following definition:

\textit{\text{Warm-starting parameters configuration:} warm-starting parameters configuration, denoted by $\psi_k$, contains values for variable parameters. These value are obtained through prior multi-objective optimisation with objective function \refeqs{object_func}:}
\begin{equation}
\label{object_func}
    U(\Omega)=(s_{\omega_1}v_{\omega_1},s_{\omega_2}v_{\omega_2},\dots,s_{\omega_{|\Omega|}} v_{\omega_{|\Omega|}})\bullet\begin{pmatrix}
        \mathbb{1}_{\omega_1}^\Omega\\\mathbb{1}_{\omega_2}^\Omega\\\dots\\\mathbb{1}_{\omega_{|\Omega|}}^\Omega
    \end{pmatrix}
\end{equation}
\textit{, where}
\begin{equation}
    \mathbb{1}_{\omega_j}^\Omega=\left\{
    \begin{aligned}
        &1, \quad \textit{if $\omega_j$ is an optimisation objective of user} \\
        &0, \quad Otherwise
    \end{aligned}\right.
\end{equation}
The warm-starting parameters configuration, denoted by $\psi_k$, comprises values for variable parameters obtained from previous multi-objective optimizations. These optimizations aim to minimize the objective function, $U(\Omega)$, representing a weighted combination of different objectives' values $v_{\omega_j}$ of the blockchain system, each weighted by a coefficient $s_{\omega_j}$ that reflects its importance in the study.

\begin{equation}
\label{objective_func}
\mathop{min}\limits_{\Phi}U(\Omega)
\end{equation}
Warm-starting parameters configuration is a Pareto efficient solution of \refeqs{objective_func}, denoted by $\psi_k$, where $\psi_k=\{c_{\phi_1}^{\psi_k}, c_{\phi_2}^{\psi_k},...,c_{\phi_n}^{\psi_k}\}$. And the set of warm-starting parameters configuration is the set of all Pareto efficient solution of \refeqs{objective_func}, denoted by $\Psi$, where $\Psi=\{\psi_1,\psi_2,\dots,\psi_{|\Psi|}\}$.

When a user interacts with GFOBS, the system first matches the selected optimization objective and retrieves the corresponding database. It then computes the Jaccard similarity between the parameter ranges of the current task and those of prior tasks. The warm-starting parameter configurations are ranked in descending order of similarity, and the top \( N \) configurations are selected as the warm-start set. Higher Jaccard similarity indicates greater alignment and a higher likelihood of faster convergence. The optimizer is initialized with this selected set, and the optimization begins from these initial values. This method accelerates convergence and enhances overall optimization efficiency. The process is summarized in pseudo-code \ref{warm-starting parameters configuration}.

\begin{algorithm}
%\textsl{}\setstretch{1.8}
\begin{scriptsize}
\floatname{algorithm}{Pseudo-code}
%\begin{algorithm} [H]
\caption{\textcolor{black}{Optimal warm-starting parameters configuration set generation}}
\label{warm-starting parameters configuration}
\begin{algorithmic}[1]
\State \textbf{Input:} The set of warm-starting parameters configuration $\Psi$
\State \textbf{Output:} Optimal warm-starting parameters configuration set $\Psi^*$
\Procedure{Select the optimal warm-starting parameters configuration}{}
    \State $\boldsymbol{a}=(\mathbb{1}_{\phi_1}^\Phi R_{\phi_1},\mathbb{1}_{\phi_2}^\Phi R_{\phi_2},\dots,\mathbb{1}_{\phi_{|\Phi|}}^\Phi R_{\phi_{|\Phi|}})$
    \For{each $\psi_k$ in $\Psi$}
        \State $\boldsymbol{b}=\begin{pmatrix}
            \mathbb{1}_{\phi_1}^\Phi R_{\phi_1}^{\psi_k}\\\mathbb{1}_{\phi_2}^\Phi R_{\phi_2}^{\psi_k}\\\dots\\\mathbb{1}_{\phi_{|\Phi|}}^\Phi R_{\phi_{|\Phi|}}^{\psi_k}
        \end{pmatrix}$
        \State $JS_{\psi_k}=\frac{|\boldsymbol{a}\cap\boldsymbol{b}|}{|\boldsymbol{a}\cup\boldsymbol{b}|}$
    \EndFor
    \State Sort the elements in $\Psi$ in an descending order of $JS_{\psi_k}$
    \State Cut the first N elements from the sorted array, denote as $\Psi^*$
    \State \textbf{Return } $\Psi^*$
\EndProcedure
\end{algorithmic}
%\end{algorithm}
\end{scriptsize}
\end{algorithm}

\subsection{Concurrent Processing (CMP)}

\begin{figure}[htp]
    \centering
    \includegraphics[width=0.8\textwidth]{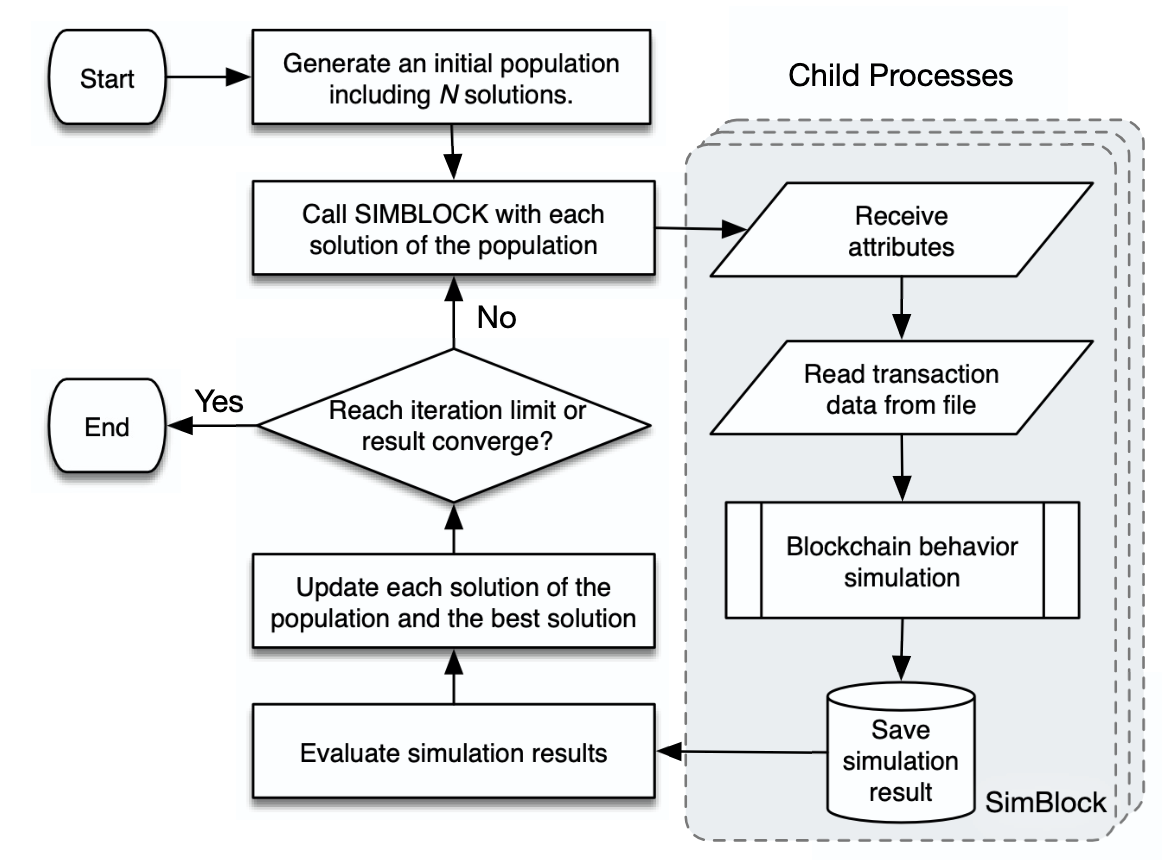}
    \caption{Schematic diagram of concurrent processing for optimization algorithms}
    \label{PSOSimulation}
\end{figure}

SIMBLOCK’s event-driven nature makes individual simulation tasks time-consuming. Running \( N \) simulations sequentially in each iteration leads to low efficiency. To address this, GFOBS implements a concurrent processing method, allowing multiple simulations to run in parallel and significantly improving performance, as illustrated in \reffig{PSOSimulation}.

This method utilizes Python’s \texttt{multiprocessing} package, which enables parallel execution by spawning separate processes—each with its own interpreter and memory space—thus bypassing Python’s Global Interpreter Lock (GIL). Unlike threading, multiprocessing is well-suited for CPU-bound tasks such as blockchain simulations. By distributing simulation tasks across multiple CPU cores, GFOBS maximizes resource utilization and achieves substantial speed-ups. This scalable approach is particularly beneficial for computationally intensive blockchain simulations, offering enhanced efficiency and adaptability to varying workloads.

\section{Experiments}\label{Exp}
Experiment 1 validated the versatility of GFOBS by applying it to the scenarios explored in a benchmark paper \cite{akbari2020impact} by Akbari \cite{akbari2020impact} in the blockchain domain, referred to hereafter as the “benchmark paper" \cite{akbari2020impact}. GFOBS automated the optimization process for block size and expected mining interval parameters, demonstrating its ability to efficiently explore the parameter space. The objective was to highlight the versatility and effectiveness of GFOBS in the process.

In Experiment 2, we conducted detailed ablation experiments on every module of GFOBS and demonstrated the role of each module. The objective was to understand the individual and combined impacts of GFOBS's two advanced techniques: warm-starting and concurrent processing.

\subsection{Experiment Settings}

Our experiments were conducted on a custom platform built using Huawei Cloud Stack (HCS), with support from the Super Intelligent Computing Center (SICC) at the University of Macau. The testbed consisted of an Ubuntu 18.04 virtual machine with 48 vCPUs at 2.50GHz, 192GB RAM, and 1TB ROM.

We evaluated three leading blockchain simulators: Bitcoin-Simulator, SIMBLOCK, and BlockSim. Their initial parameter settings are provided in Table \ref{blockchain simulator parameters}. Parameters not shared across all simulators retain their default values. Common parameters include block size, expected mining interval, average hash rate, node count, region distribution, latency, and upload/download bandwidth. These were unified across simulators to minimize inconsistencies, as detailed in Table \ref{blockchain simulator parameters}. Simulator-specific parameters remained at their defaults.

\begin{figure}[htbp]
\centering

\begin{minipage}[t]{0.48\textwidth}
\centering
\includegraphics[width=\textwidth]{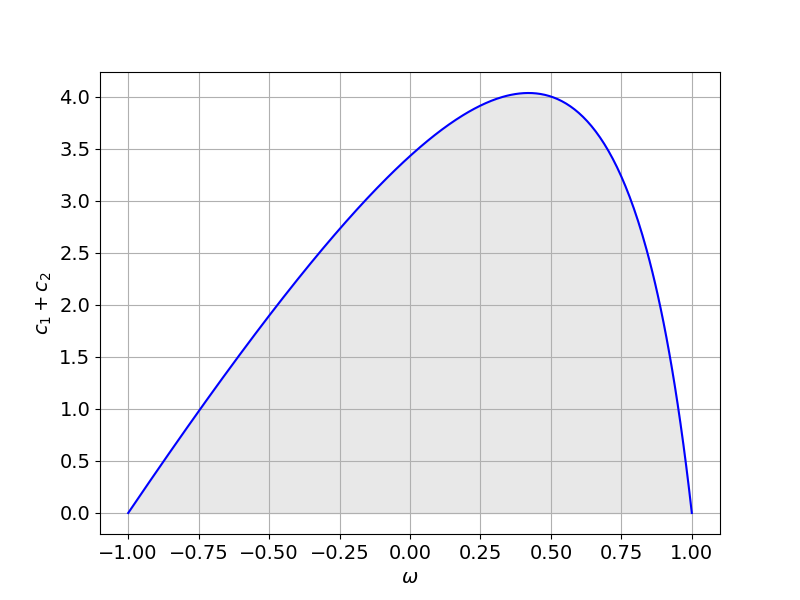}
\end{minipage}

\caption{Inequality bewteen $\omega$ and $c_1$+$c_2$}
\label{inequality_plot}
\end{figure}

Optimization algorithm settings are summarized in Table \ref{tab:optimization_algorithms}, with the following rationale: (i) All algorithms used a population size of 50 and 100 iterations, resulting in 5000 total evaluations, balancing efficiency and solution quality. The lower and upper bounds for variable parameters were set to [1s, 1KB] and [1800s, 50MB], respectively. (ii) For Genetic Algorithms (GA), the mutation probability was set to 0.001 to avoid random search behavior, as recommended in \cite{lu2004incremental}. (iii) For Differential Evolution (DE), we followed recommendations from \cite{mallipeddi2011differential}, setting $F=0.5$ and $\text{CR}=0.3$. These values ensure sufficient diversity and convergence speed without excessive disturbance. (iv) For Particle Swarm Optimization (PSO), based on the convergence condition in \cite{poli2007exact,poli2009mean,harrison2017optimal}:
\begin{equation}
    \label{pso3}
    c_1 + c_2 < \frac{24(1 - \omega^2)}{7 - 5\omega},
\end{equation}
we chose $\omega = 0.5$, $c_1 = 0.8$, and $c_2 = 0.8$ to ensure stable convergence, as illustrated in Figure \ref{inequality_plot}.

\subsection{Benchmarks}
In this study, we utilize time to convergence and average computational resource utilization as benchmarks for evaluating the efficiency of GFOBS.
\begin{itemize}
    
    \item \text{Time to convergence (ToC):} Measure the time taken for the optimization process to converge to a solution. Compare this time with and without the use of GFOBS's advanced features like warm-starting and concurrent processing. A significant reduction in convergence time indicates higher efficiency.
    
    %In this study, besides throughput, we also utilize forking rate as benchmarks for evaluating the scalability and stability of the Blockchain network.
    
    \item \text{Average Computational resource utilization (ACRU):} Monitor the resources (CPU usage) consumed during the optimization process and take its average at the end of optimization process. Efficient use of computational resources, especially in simulations using concurrent processing, reflects higher efficiency.
    
\end{itemize}

\subsection{Experiment 1: Enhancing Simulation-Based Blockchain Research with GFOBS}

Akbari et al. \cite{akbari2020impact} investigated the impact of block size and mining interval on fork rate in blockchain networks. Their simulation-based study aimed to identify parameter combinations that maintain a fork rate below 10\%, testing six block sizes (1 KB to 25 MB) and eight mining intervals (1s to 1800s), resulting in 48 scenarios. Each simulation, run with 100 blocks, took ~70 seconds, totaling 3,360 seconds. They found that a 25 MB block size with a 600-second interval achieved the highest throughput (83 TPS) under the fork rate threshold.

To evaluate the effectiveness of GFOBS, we replicate this experiment using Akbari’s methodology as a baseline. This enables a direct comparison and highlights the enhancements brought by GFOBS. We also showcase GFOBS's interoperability by integrating multiple blockchain simulators within the framework. While Akbari’s approach offered valuable insights, it relied on manual exploration and did not guarantee optimal parameter discovery. In contrast, GFOBS automates this process by iteratively optimizing block size and mining interval to efficiently identify the most effective configuration. This not only accelerates the optimization process but also reduces manual workload, enhancing the practicality of simulation-based blockchain research.

\subsubsection{Optimizing Block Size and Expected Mining Interval for Fork Rate and Throughput}
To build upon and enhance the experiment conducted by Akbari et al., we applied GFOBS to the same research issue. Focusing on optimizing the block size and expected mining interval parameters, we aimed to improve both the fork rate and throughput. The optimization algorithm implemented is PSO, chosen for its lower Time to Convergence (ToC) compared to other optimization algorithms evaluated in Experiment 2. The results of this optimization, utilizing different blockchain simulators, are presented in Table \ref{optimization_res}.

\begin{table}
\caption{Optimization Results with GA, DE and PSO}
\label{optimization_res}
\centering
%\resizebox{\linewidth}{!}{
\begin{tabular}{p{3cm}<{\raggedright} p{1.8cm}<{\raggedright}  p{1.8cm}<{\raggedright}  p{1.8cm}<{\raggedright}  p{1.8cm}<{\raggedright}}
\hline
\hline
\textbf{Algorithm} & \textbf{Block size} & \textbf{Expected mining interval} & \textbf{Throughput} & \textbf{Fork rate} \\ 
\hline
\multirow{1}{*}{BlockSim}  & 24.8MB & 603s & 95 & 9.64\% \\ 

\multirow{1}{*}{\textbf{SIMBLOCK}}  & \textbf{27.4MB} & \textbf{612s} & \textbf{103} & \textbf{9.98\%} \\ 

\multirow{1}{*}{Bitcoin-Simulator} & 25.2MB & 607s & 94 & 9.25\% \\ 
\hline
\textbf{Benchmark paper \cite{akbari2020impact}} & 25MB & 600s & 83 & $\leq$10\% \\
\hline
\end{tabular}%}
\end{table}

\subsubsection{Comparative Analysis of Benchmark paper \cite{akbari2020impact} and GFOBS Optimization}
This section delves into the comparative analysis between the findings from \cite{akbari2020impact} and the outcomes achieved through our GFOBS optimization. The optimization results, displayed in Table \ref{optimization_res}, are derived from three different blockchain simulators, compared with the boundary conditions identified in the benchmark study. Notably, the SIMBLOCK simulator, when employed within GFOBS, delineated the most precise boundary for block size and expected mining interval, optimizing close to the critical fork rate threshold of 10\%.

\cite{akbari2020impact} established a boundary where the fork rate stays below 10\%, providing a safe operational zone. For instance, they suggested that a 25MB block size with a 600-second mining interval resulted in an acceptable fork rate under this threshold. However, this does not ascertain the exact point where the fork rate approaches the maximum permissible limit, which could lead to optimal network utilization.

In contrast, our GFOBS framework refines this approach, achieving a fork rate of exactly 9.98\% with a 27.4MB block size and 612-second mining interval. This outcome demonstrates the ability of GFOBS to precisely calibrate the blockchain parameters, thus identifying a "precisest boundary." This term signifies the specific parameter combination that allows the blockchain network to operate at maximum efficiency, pushing the fork rate to as close to 10\% as possible without exceeding it, ensuring optimal throughput and resource utilization.

The results obtained through GFOBS not only exhibited a higher throughput of 103 TPS but also maintained the fork rate within the narrowly defined optimal threshold, showcasing an improved level of precision and efficiency compared to the benchmark study. The ability of GFOBS to achieve such targeted results confirms the effectiveness of its optimization process and its advantage in facilitating more accurate and efficient blockchain simulations.

\textcolor{black}{Additionally, the effective application of GFOBS with three different blockchain simulators highlights its capability for broad compatibility. This attribute allows for straightforward integration with various simulation tools, confirming the flexibility and reliability of its interface in diverse simulation settings.}

\subsection{Experiment 2: Ablation study of GFOBS's modules}
Experiment 2 is structured around 12 distinct optimization tests, as shown in TABLE \ref{ablation}, to evaluate the efficacy of the GFOBS framework. In all optimization tests, we are going to solve the same optimization problem in \cite{akbari2020impact}. Therefore, the optimisation parameters are block size and expected mining interval, and the optimisation objectives are throughput and fork rate. Additionally, since SIMBLOCK outperforms other blockchain simulators in experiment 1, we use SIMBLOCK as the simulator component of GFOBS in experiment 2.

\begin{table*}[]
\footnotesize
\centering
\caption{\textcolor{black}{Ablation study of GFOBS's modules}}
\label{ablation}
\begin{threeparttable}
\begin{tabular}{p{0.6cm}<{\raggedright} | p{0.5cm}<{\raggedright} | p{0.5cm}<{\raggedright} | p{1cm}<{\raggedright} | p{1cm}<{\raggedright} | p{2cm}<{\raggedright} | p{1cm}<{\raggedright} | p{1.3cm}<{\raggedright} | p{0.8cm}<{\raggedright}}
\hline
\hline
Group              & WS                                          & CMP                                         & Average ToC$\downarrow$               & Average ACRU$\uparrow$             & Optimization Test & Algorithm & ToC       & ACRU    \\ \hline
\multirow{3}{*}{1} & \multirow{3}{*}{\ding{56}} & \multirow{3}{*}{\ding{56}} & \multirow{3}{*}{6388.48s} & \multirow{3}{*}{4.36\%}  & test 1a                 & DE        & 10907.82s & 2.47\%  \\ \cline{6-9} 
                   &                                             &                                             &                           &                          &test 2a                 & GA        & 7169.34s  & 5.23\%  \\ \cline{6-9} 
                   &                                             &                                             &                           &                          &test 3a                 & PSO       & 1083.28s  & 5.37\%  \\ \hline
\multirow{3}{*}{2} & \multirow{3}{*}{\ding{51}} & \multirow{3}{*}{\ding{56}} & \multirow{3}{*}{3566.69s} & \multirow{3}{*}{4.38\%}  &test 1b                 & DE        & 6391.09s  & 2.47\%  \\ \cline{6-9} 
                   &                                             &                                             &                           &                          &test 2b                 & GA        & 3802.89s  & 5.32\%  \\ \cline{6-9} 
                   &                                             &                                             &                           &                          &test 3b                & PSO       & 506.08s   & 5.34\%  \\ \hline
\multirow{3}{*}{3} & \multirow{3}{*}{\ding{56}} & \multirow{3}{*}{\ding{51}} & \multirow{3}{*}{3198.76s} & \multirow{3}{*}{43.13\%} &test 1c                 & DE        & 6026.58s  & 44.35\% \\ \cline{6-9} 
                   &                                             &                                             &                           &                          &test 2c                 & GA        & 2875.41s  & 44.06\% \\ \cline{6-9} 
                   &                                             &                                             &                           &                          &test 3c                & PSO       & 694.29s   & 40.98\% \\ \hline
\multirow{3}{*}{4} & \multirow{3}{*}{\ding{51}} & \multirow{3}{*}{\ding{51}} & \multirow{3}{*}{1826.71s} & \multirow{3}{*}{43.48\%} &test 1d                 & DE        & 3282.58s  & 44.58\% \\ \cline{6-9} 
                   &                                             &                                             &                           &                          &test 2d                 & GA        & 1879.21s  & 44.50\% \\ \cline{6-9} 
                   &                                             &                                             &                           &                          &test 3d                & PSO       & 318.33s   & 41.35\% \\ \hline
\end{tabular}
\begin{tablenotes}
\footnotesize
\item WS: Warm Starting; CMP: Concurrent processing; ToC: Time to Convergence; ACRU: Average Computational Resource Utilization;
\end{tablenotes}

\end{threeparttable}
\end{table*}

We divide these optimization tests into 4 groups. Within each group, the optimization tests employ different optimization algorithms such as  DE, GA, PSO while utilizing the same modules of GFOBS. We denote optimization test (1a, 2a, 3a) by Group A, optimization test (1b, 2b, 3b) by Group B, optimization test (1c, 2c, 3c) by Group C and optimization test (1d, 2d, 3d) by Group D.

The first group of optimization tests serves as a baseline, where we address the optimization of SIMBLOCK's parameters without employing either the warm-starting technique or the concurrent processing method. The second and third groups of optimization tests introduce these techniques individually. Specifically, the second group of optimization tests incorporates the warm-starting technique, while the third group of optimization tests employs the concurrent processing technique. These optimization tests are designed to assess the individual impacts of each technique and will be compared with the first group of optimization tests to highlight their respective advantages. Finally, the fourth group of optimization tests represents the full implementation of GFOBS, integrating both the warm-starting technique and the concurrent processing technique. This comprehensive approach is expected to demonstrate significant enhancements in optimization efficiency and is contrasted against the baseline simulation to underscore the combined benefits of these advanced techniques.

\begin{figure}[htbp]
\centering

\begin{minipage}[t]{0.8\textwidth}
\centering
\includegraphics[width=\textwidth]{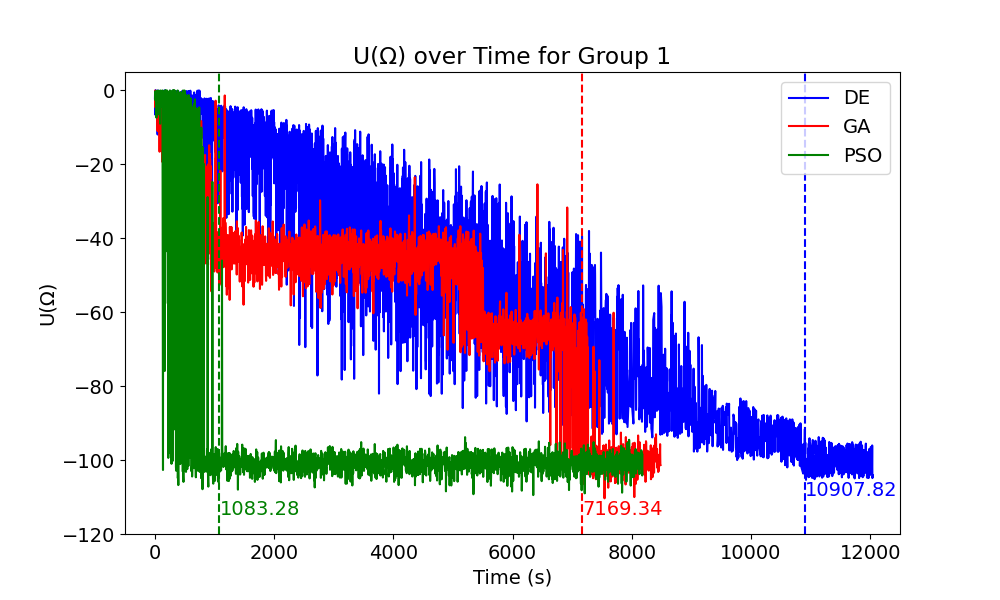}
\end{minipage}

\begin{minipage}[t]{0.8\textwidth}
\centering
\includegraphics[width=\textwidth]{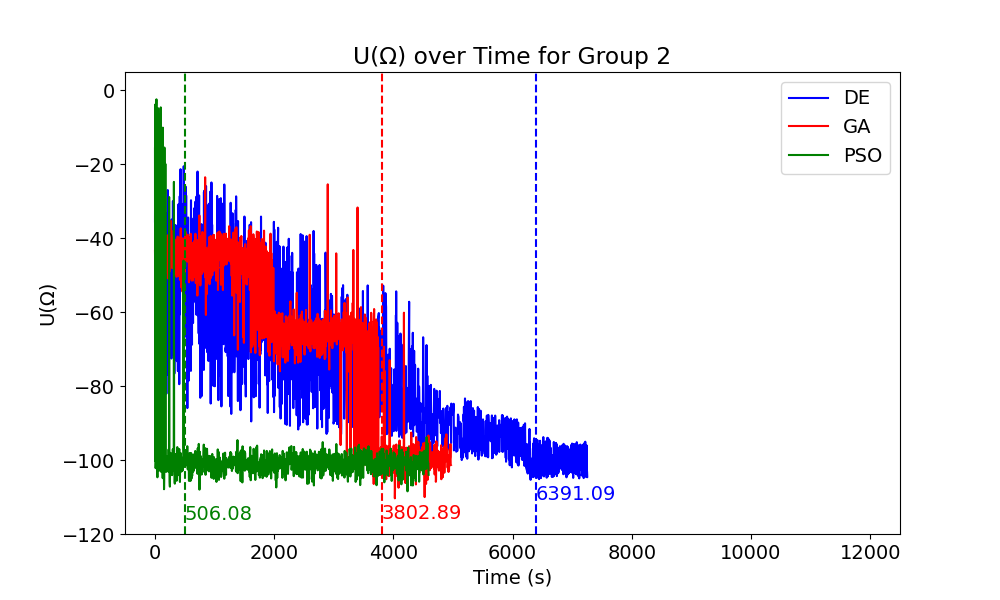}
\end{minipage}

\begin{minipage}[t]{0.8\textwidth}
\centering
\includegraphics[width=\textwidth]{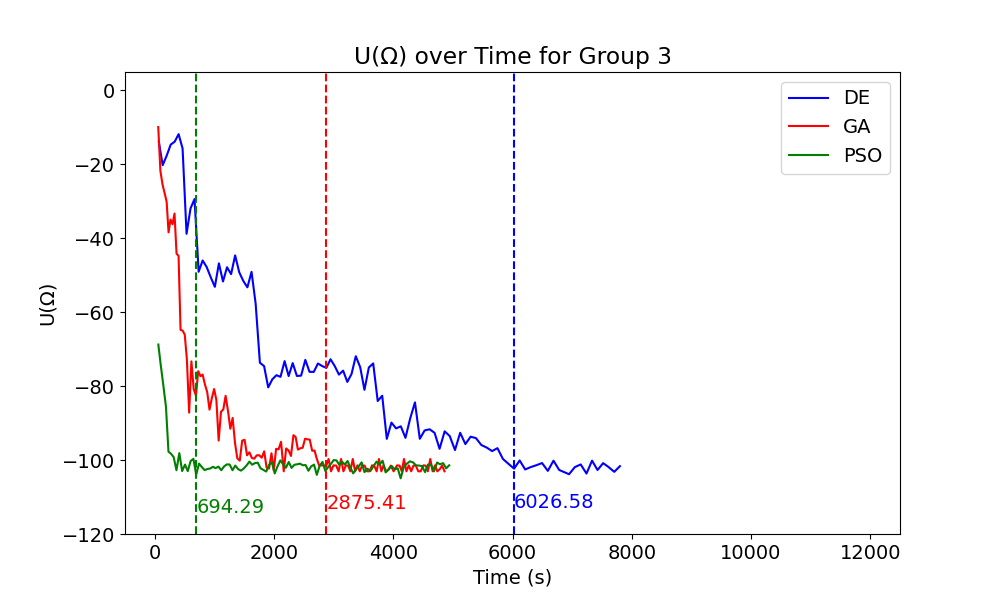}
\end{minipage}

\begin{minipage}[t]{0.8\textwidth}
\centering
\includegraphics[width=\textwidth]{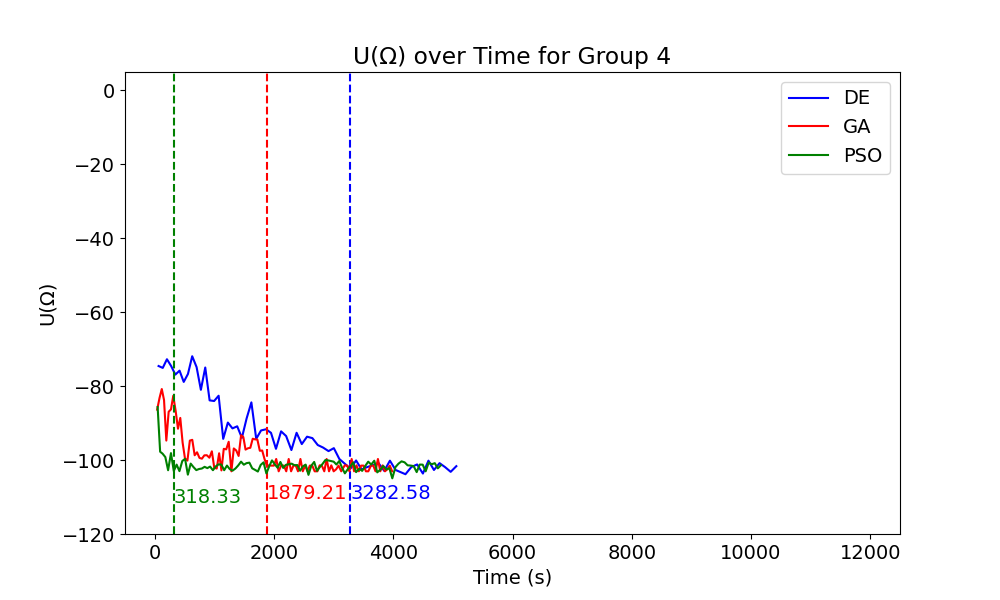}
\end{minipage}

\caption{Time to Convergence of all groups}
\label{ToC_result}
\end{figure}

The ToC obtained in each group of optimization tests are shown in \reffig{ToC_result}. If the entire population (consisting of all individuals) attains an identical state or solution during a certain iteration, we say that the algorithm has converged at that point. The dotted line represents the ToC.

\begin{figure}[htbp]
\centering

\begin{minipage}[t]{0.8\textwidth}
\centering
\includegraphics[width=\textwidth]{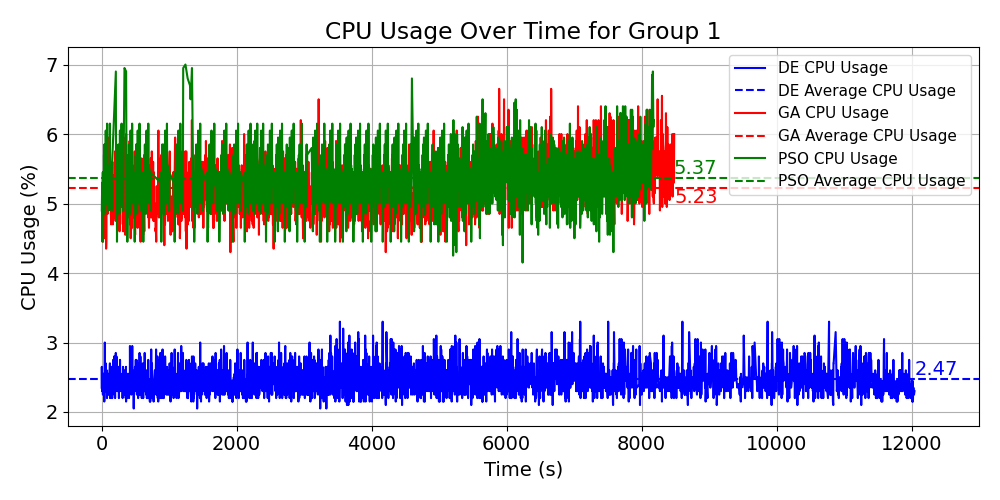}
\end{minipage}

\begin{minipage}[t]{0.8\textwidth}
\centering
\includegraphics[width=\textwidth]{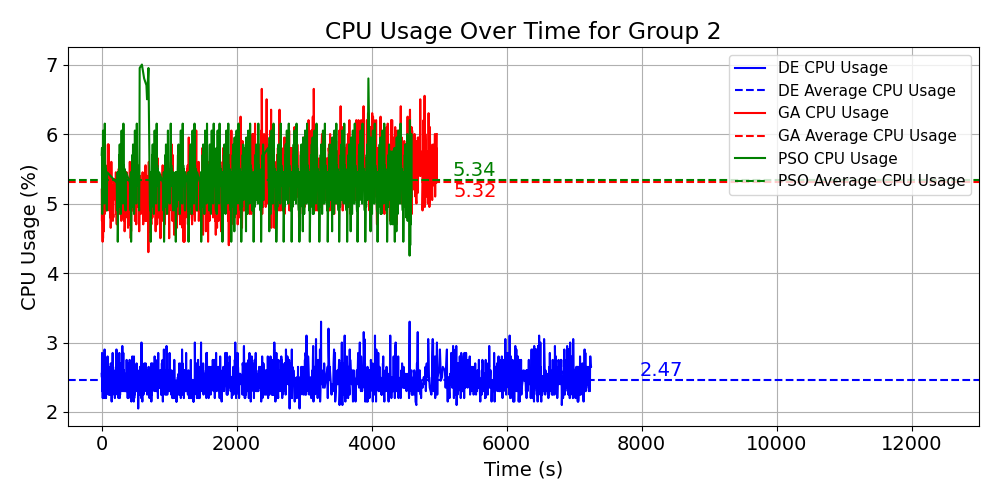}
\end{minipage}

\begin{minipage}[t]{0.8\textwidth}
\centering
\includegraphics[width=\textwidth]{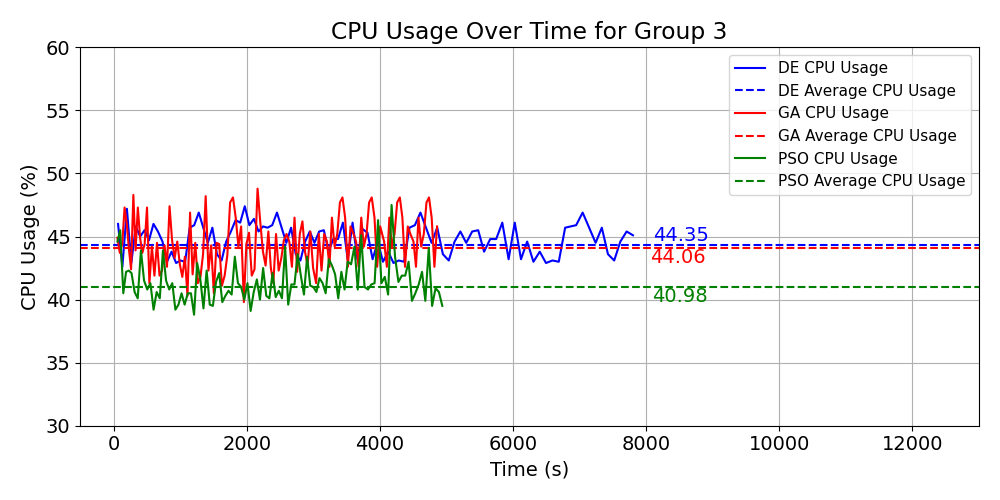}
\end{minipage}

\begin{minipage}[t]{0.8\textwidth}
\centering
\includegraphics[width=\textwidth]{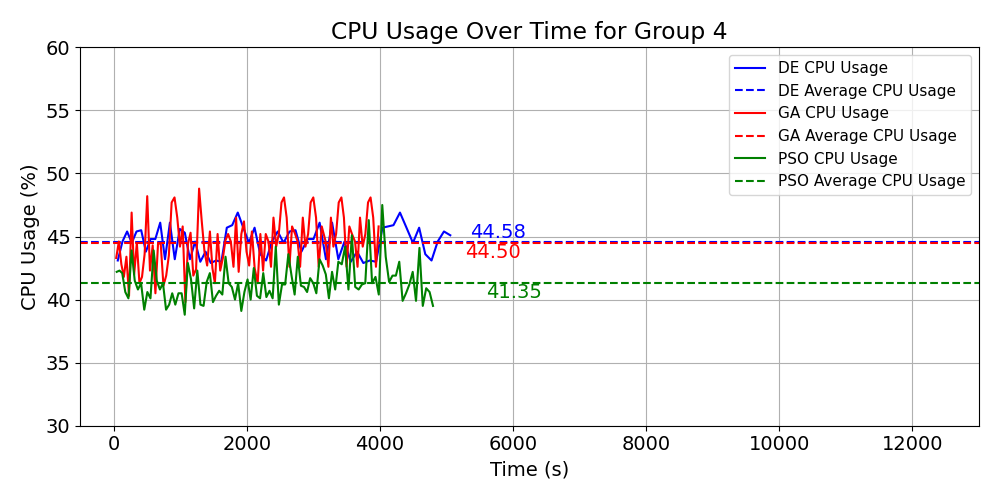}
\end{minipage}

\caption{Computational Resource Utilization of all groups}
\label{ACRU_result}

\end{figure}

The ACRU obtained in each group of optimization tests are shown in \reffig{ACRU_result}, which reflects the CPU usage throughout the optimisation process. The dotted line represents the ACRU, which is the average of CPU usage.

\subsubsection{\textcolor{black}{Group A: Optimisation without Warm-Starting and Concurrent processing}}
In Group A, the optimization challenge is addressed using conventional methods external to the GFOBS framework. The initial values for the optimization are selected randomly, lacking the strategic advantage that warm-starting provides. This randomness can lead to starting the optimization process in less promising areas of the solution space, potentially prolonging the search for optimal solutions. Additionally, the execution of blockchain simulation tasks follows a sequential pattern, further contributing to inefficiencies. The ToC for this group is considerably high, with an average of 6388.48 seconds, reflecting the extended duration required to reach optimization under this traditional approach. The ACRU is moderately set at 4.36\%, indicating that the computational resources are not fully leveraged, possibly due to the sequential nature of task execution which fails to utilize the available processing power effectively.

\subsubsection{\textcolor{black}{Group B: Optimization with Warm-Starting only}}
Group B incorporates the warm-starting technique from the GFOBS, where initial values for the simulation are informed by the outcomes of previous optimizations. This method significantly reduces the average ToC by 44.17\%, underscoring the effectiveness of warm-starting in expediting the convergence process. However, the ACRU in this group remains moderate at 4.38\%, similar to Group A. This observation suggests that while warm-starting accelerates the optimization by leveraging historical data, the overall resource utilization does not markedly improve, possibly because the sequential processing of tasks still prevails.

\subsubsection{\textcolor{black}{Group C: Optimization with Concurrent processing only}}
Group C explores the impact of employing concurrent processing within the GFOBS framework, excluding the warm-starting feature. This approach markedly decreases the average ToC by 49.93\% compared to Group A, showcasing the significant efficiency gains from processing multiple simulation tasks in parallel. Notably, the ACRU experiences a substantial increase, surging by 889.22\%. This dramatic rise indicates the enhanced capability of concurrent processing to maximize computational resource utilization, drastically improving the efficiency of the optimization process.

\subsubsection{\textcolor{black}{Group D: Optimisation with Warm-Starting and Concurrent processing}}
In Group D, the optimization process within GFOBS integrates both warm-starting and concurrent processing techniques, harnessing the full potential of the framework. This combination results in the most significant reduction in average ToC, plummeting by 71.41\% relative to Group A, and illustrating the profound impact of combining these two optimization enhancements. The ACRU for this group is the highest among all, at 43.48\%, indicating an optimal utilization of computational resources. This high level of resource utilization reflects the synergistic effect of warm-starting and concurrent processing in streamlining the optimization process, thereby achieving the fastest convergence and the most efficient use of computational resources in the optimization of blockchain simulations.

\subsubsection{\textcolor{black}{Comparative analysis}}
In our comparative analysis, we delve into the performance metrics across the different groups by evaluating the average ToC and average ACRU. These metrics are crucial in understanding the impact of each optimization strategy employed within the GFOBS framework. The summarized results in Table \ref{ablation} provide a comprehensive overview of how each group fared in the optimization tests.

The analysis indicates a distinct hierarchy in the performance of the optimization strategies across the groups. Group D stands out with the most impressive results, achieving the shortest average ToC of 1826.71 seconds and the highest ACRU at 43.48\%. This demonstrates the combined strength of warm-starting and concurrent processing in optimizing blockchain simulations, where the integration of these techniques significantly accelerates the convergence time and maximizes computational resource utilization.

Group C, which employs only concurrent processing, also demonstrates a notable performance, underscoring the impact of parallel task execution in reducing optimization time. Although it does not reach the efficiency levels of Group D, it substantially surpasses Group B, which utilizes only warm-starting. This finding highlights the substantial benefit of concurrent processing in enhancing the speed of the optimization process.

Group B, relying solely on the warm-starting technique, shows improved efficiency over Group A, which adheres to traditional optimization methods without any enhancements. While warm-starting alone accelerates the optimization process by utilizing insights from previous iterations, the absence of concurrent processing limits its potential to optimize resource usage fully.

The descending trend in ToC from Group D to Group A, paired with the ascending trend in ACRU, starkly illustrates the incremental benefits of incorporating warm-starting and concurrent processing into the optimization framework. Each addition or integration of these advanced methods brings a substantial improvement in both the speed and resource efficiency of the optimization process.

\section{Conclusion}
In this study, we introduced the Generic Framework for Optimization in Blockchain Simulators (GFOBS), a comprehensive and adaptable tool crafted to refine and enhance the optimization processes within blockchain simulations. GFOBS is engineered to navigate the intricate challenges of blockchain technology, integrating a spectrum of optimization algorithms and supporting an extensive array of optimization variables and objectives. This integration is pivotal in addressing the multifaceted nature of blockchain-related issues.

The core of GFOBS lies in its modular design, which permits seamless incorporation of various blockchain simulators and optimization algorithms, offering unparalleled flexibility and customization options. This design is complemented by the warm-starting technique, which leverages previous optimization outcomes to inform and accelerate subsequent simulations, thus streamlining the optimization journey and ensuring more efficient use of computational resources. Furthermore, the concurrent processing feature within GFOBS underscores its capacity to handle multiple simulation tasks in parallel, drastically reducing the time required for comprehensive simulation studies and enhancing the overall efficiency of the simulation process.

Despite the innovative features and capabilities of GFOBS, it is not without its limitations. The framework's effectiveness is somewhat dependent on the selection of blockchain simulators and optimization algorithms, which might limit its applicability across a broader spectrum of blockchain simulation scenarios. The current setup of GFOBS may not fully embrace the rapid advancements in optimization methods and blockchain simulation technologies, indicating room for further enhancement and expansion.

There is substantial scope for augmenting the GFOBS framework. The integration of advanced optimization algorithms, particularly those stemming from the realms of machine learning and artificial intelligence, could significantly bolster the framework's optimization prowess. These sophisticated algorithms promise to refine the optimization process, ensuring more precise and efficient outcomes. Moreover, broadening the compatibility of GFOBS to encompass a wider array of blockchain simulators would enhance its versatility and applicability, making it a more robust and comprehensive tool for blockchain simulation optimization.

Furthermore, the scalability and complexity of blockchain networks are anticipated to escalate, necessitating a corresponding evolution in GFOBS to adequately meet these advancing demands. Ensuring that GFOBS can effectively manage and optimize simulations for larger and more complex blockchain networks will be crucial for maintaining its relevance and efficacy in the field.

%\vspace{-0.2cm}
\section*{Acknowledgments}
This research was funded by the University of Macau (file no. MYRG2022-00162-FST and MYRG2019-00136-FST).

\end{document}